# A Matlab Program to Calculate the Maximum Entropy Distributions

Ali Mohammad-Djafari

*Laboratoire des Signaux et Systèmes,
Supélec, Plateau de Moulon, 91192 Gif-sur-Yvette, France*

## INTRODUCTION

Shannon (1948) indicated how maximum entropy (ME) distributions can be derived by a straigtforward application of the calculus of variations technique. He defined the entropy of a probability density function $p(x)$ as

$$H = -\int p(x) \ln p(x) \, \mathrm{d}x \tag{1}$$

Maximizing $H$ subject to various side conditions is well–known in the literature as a method for deriving the forms of minimal information prior distributions; e.g. Jaynes (1968) and Zellner (1977). Jaynes (1982) has extensively analyzed examples in the discrete case, while in Lisman and Van Znylen (1972), Rao (1973) and Gokhale (1975), Kagan, Linjik continuous cases are considered. In the last case, the problem, in its general form, is the following

$$\begin{aligned}
\text{maximize} \quad H &= -\int p(x) \ln p(x) \, \mathrm{d}x \\
\text{subject to} \quad \mathrm{E}\{\phi_n(x)\} &= \int \phi_n(x) p(x) \, \mathrm{d}x = \mu_n, \quad n = 0, \ldots, N
\end{aligned} \tag{2}$$

where $\mu_0 = 1$, $\phi_0(x) = 1$ and $\phi_n(x), n = 0, \ldots, N$ are $N$ known functions, and $\mu_n, n = 0, \ldots, N$ are the given expectation data. The classical solution of this problem is given by

$$p(x) = \exp\left[-\sum_{n=0}^{N} \lambda_n \phi_n(x)\right] \tag{3}$$

The $(N+1)$ Lagrangien parameters $\boldsymbol{\lambda} = [\lambda_0, \ldots, \lambda_n]$ are obtained by solving the following set of $(N+1)$ nonlinear equations

$$G_n(\boldsymbol{\lambda}) = \int \phi_n(x) \exp\left[-\sum_{n=0}^{N} \lambda_n \phi_n(x)\right] \mathrm{d}x = \mu_n, \quad n = 0, \ldots, N \tag{4}$$

The distributions defined by (3) form a great number of known distributions which are obtained by choosing the appropriate $N$ and $\phi_n(x), n = 0, \ldots, N$. In general $\phi_n(x)$ are either the powers of $x$ or the logarithm of $x$. See Mukhrejee and Hurst (1984), Zellner (1988), Mohammad–Djafari (1990) for many other examples and discussions. Special cases have been extensively analyzed and used by many authors. When $\phi_n(x) = x^n, n = 0, \ldots, N$ then $\mu_n, n = 0, \ldots, N$ are the given $N$ moments of the distribution. See, for example, Zellner (1988) for a numerical implementation in the case $N = 4$.

In this communication we propose three programs written in MATLAB to solve the system of equations (4). The first is a general program where $\phi_n(x)$ can be any functions. The second is a special case where $\phi_n(x) = x^n, n = 0, \ldots, N$. In this case the $\mu_n$ are the geometrical moments of $p(x)$. The third is a special case where $\phi_n(x) = \exp(-jn\omega x), n = 0, \ldots, N$. In this case the $\mu_n$ are the trigonometrical moments (Fourier components) of $p(x)$. We give also some examples to illustrate the usefullness of these programs.

## PRINCIPLE OF THE METHOD

We have seen that the solution of the standard ME problem is given by (3) in which the Lagrange multipliers $\boldsymbol{\lambda}$ are obtained by solving the nonlinear equations (4). In general, these equations are solved by the standard Newton method which consists of expanding $G_n(\boldsymbol{\lambda})$ in Taylor's series around trial values of the *lambda*'s, drop the quadratic and higher order terms, and solve the resulting linear system iteratively. We give here the details of the numerical method that we implemented. When developing the $G_n(\boldsymbol{\lambda})$ in equations (4) in first order Taylor's series around the trial $\boldsymbol{\lambda}^0$, the resulting linear equations are given by

$$G_n(\boldsymbol{\lambda}) \cong G_n(\boldsymbol{\lambda}^0) + (\boldsymbol{\lambda} - \boldsymbol{\lambda}^0)^t \left[\mathbf{grad}\, G_n(\boldsymbol{\lambda})\right]_{(\boldsymbol{\lambda}=\boldsymbol{\lambda}^0)} = \mu_n, \quad n = 0, \ldots, N \quad (5)$$

Noting the vectors $\boldsymbol{\delta}$ and $\boldsymbol{v}$ by

$$\boldsymbol{\delta} = \boldsymbol{\lambda} - \boldsymbol{\lambda}^0$$

$$\boldsymbol{v} = \left[\mu_0 - G_0(\boldsymbol{\lambda}^0), \ldots, \mu_N - G_N(\boldsymbol{\lambda}^0)\right]^t$$

and the matrix $\boldsymbol{G}$ by

$$\boldsymbol{G} = \left[\, g_{nk} \,\right] = \left[ \frac{\partial G_n(\boldsymbol{\lambda})}{\partial \lambda_k} \right]_{(\boldsymbol{\lambda}=\boldsymbol{\lambda}^0)} \quad n, k = 0, \ldots, N \quad (6)$$

then equations (5) become

$$\boldsymbol{G}\boldsymbol{\delta} = \boldsymbol{v} \quad (7)$$

This system is solved for $\boldsymbol{\delta}$ from which we drive $\boldsymbol{\lambda} = \boldsymbol{\lambda}^0 + \boldsymbol{\delta}$, which becomes our new initial vector $\boldsymbol{\lambda}^0$ and the iterations continue until $\boldsymbol{\delta}$ becomes appropriately small. Note that the matrix $\boldsymbol{G}$ is a symmetric one and we have

$$g_{nk} = g_{kn} = -\int \phi_n(x)\,\phi_k(x)\,\exp\left[-\sum_{n=0}^{N}\lambda_n\,\phi_n(x)\right]\,\mathrm{d}x \quad n,k=0,\ldots,N \quad (8)$$

So in each iteration we have to calculate the $N(N-1)/2$ integrals in the equation (8). The algorithm of the general Maximum Entropy problem is then as follows:

1. Define the range and the discretization step of $x$             (xmin, xmax,dx).
2. Write a function to calculate $\phi_n(x), n=0,\ldots,N$             (fin_x).
3. Start the iterative procedure with an initial estimate $\boldsymbol{\lambda}^0$             (lambda0).
4. Calculate the $(N+1)$ integrals in equations (4) and the $N(N-1)/2$ distinct elements $g_{nk}$ of the matrix $\boldsymbol{G}$ by calculating the integrals in the equations(8)             (Gn, gnk).
5. Solve the equation (7) to find $\boldsymbol{\delta}$             (delta).
6. Calculate $\boldsymbol{\lambda} = \boldsymbol{\lambda}^0 + \boldsymbol{\delta}$ and go back to step 3 until $\boldsymbol{\delta}$ becomes negligible.

The calculus of the integrals in equations (4) and (8) can be made by a univariate Simpson's method. We have used a very simplified version of this method.

## Case of geometrical moments

Now consider the special case of moments problem where $\phi_n(x) = x^n$, $n = 0,\ldots,N$. In this case equations (3), (4) and (8) become

$$p(x) = \exp\left[-\sum_{m=0}^{N}\lambda_m\,x^m\right] \quad (9)$$

$$G_n(\boldsymbol{\lambda}) = \int x^n \exp\left[-\sum_{m=0}^{N}\lambda_m\,x^m\right]\,\mathrm{d}x = \mu_n, \quad n=0,\ldots,N \quad (10)$$

$$g_{nk} = g_{kn} = -\int x^n x^k \exp\left[-\sum_{m=0}^{N}\lambda_m\,x^m\right]\,\mathrm{d}x = -G_{n+k}(\boldsymbol{\lambda}) \quad n,k=0,\ldots,N \quad (11)$$

This means that the $[(N+1)\times(N+1)]$ matrix $\boldsymbol{G}$ in equation (7) becomes a symmetric Hankel matrix which is entirely defined by $2N+1$ values $G_n(\boldsymbol{\lambda}), n=0,\ldots,2N$. So the algorithm in this case is the same as in the precedent one with two simplifications

1. In step 2 we do not need to write a seperate function to calculate the functions $\phi_n(x) = x^n, n=0,\ldots,N$.

2. In step 4 the number of integral evaluations is reduced, because the elements $g_{nk}$ of the matrix $\boldsymbol{G}$ are related to the integrals $G_n(\boldsymbol{\lambda})$ in equations (10). This matrix is defined entirely by only $2N+1$ components.

## Case of trigonometrical moments

Another interesting special case is the case where the data are the Fourier components of $p(x)$

$$\mathrm{E}\{\exp(-jn\omega_0 x)\} = \int \exp(-jn\omega_0 x)\, p(x)\, \mathrm{d}x = \mu_n, \quad n=0,\ldots,N, \qquad (12)$$

where $\mu_n$ may be complex–valued and has the property $\mu_{-n} = \overline{\mu_n}$. This means that we have the following relations

$$\phi_n(x) = \exp(-jn\omega_0 x), \quad n=-N,\ldots,0,\ldots N, \qquad (13)$$

$$p(x) = \exp\left[-\mathrm{Real}\sum_{n=0}^{N}\lambda_n \exp(-jn\omega_0 x)\right], \qquad (14)$$

$$G_n(\boldsymbol{\lambda}) = \int \exp(-jn\omega_0 x)\, p(x)\, \mathrm{d}x, \quad n=0,\ldots,N, \qquad (15)$$

$$g_{nk} = \begin{cases} -G_{n-k}(\boldsymbol{\lambda}) & \text{for } n \geq k, \\ -G_{n+k}(\boldsymbol{\lambda}) & \text{for } n < k \end{cases} \quad n,k=0,\ldots,N, \qquad (16)$$

so that all the elements of the matrix $\boldsymbol{G}$ are related to the discrete Fourier transforms of $p(x)$. Note that $\boldsymbol{G}$ is a Hermitian Toeplitz matrix.

## EXAMPLES AND NUMERICAL EXPERIMENTS

To illustrate the usefullness of the proposed programs we consider first the case of the Gamma distribution

$$p(x;\alpha,\beta) = \frac{\beta^{(1-\alpha)}}{\Gamma(1-\alpha)}\, x^\alpha \exp(-\beta x), \quad x>0, \alpha<1, \beta>0. \qquad (17)$$

This distribution can be considered as a ME distribution when the constraints are

$$\begin{cases} \int p(x;\alpha,\beta)\,\mathrm{d}x = 1 \\ \int x\, p(x;\alpha,\beta)\,\mathrm{d}x = \mu_1 \\ \int \ln(x)\, p(x;\alpha,\beta)\,\mathrm{d}x = \mu_2 \end{cases} \quad \begin{cases} \text{normalization} \quad \phi_0(x) = 1, \\ \phi_1(x) = x, \\ \phi_2(x) = \ln(x). \end{cases} \qquad (18)$$

This is easy to verify because the equation (12) can be written as

$$p(x;\alpha,\beta) = \exp\left[-\lambda_0 - \lambda_1 x - \lambda_2 \ln(x)\right]$$

$$\text{with} \quad \lambda_0 = -\ln \frac{\beta^{(1-\alpha)}}{\Gamma(1-\alpha)}, \quad \lambda_1 = \beta \quad \text{and} \quad \lambda_2 = -\alpha.$$

Now consider the following problem

$$\text{Given } \mu_1 \text{ and } \mu_2 \text{ determine } \lambda_0, \lambda_1 \text{ and } \lambda_2.$$

This can be done by the standard ME method. To do this, first we must define the range of $x$, (xmin, xmax, dx), and write a function fin_x to calculate the functions $\phi_0(x) = 1$, $\phi_1(x) = x$ and $\phi_2(x) = \ln x$ (See the function fin1_x in Annex). Then we must define an initial estimate $\boldsymbol{\lambda}^0$ for $\boldsymbol{\lambda}$ and, finally, let the program works.

The case of the *Gamma* distribution is interesting because there is an analytic relation between $(\alpha, \beta)$ and the mean $m = \mathrm{E}\{x\}$ and variance $\sigma^2 = \mathrm{E}\{(x-m)^2\}$ which is

$$\begin{cases} m = (1-\alpha)/\beta \\ \sigma^2 = (1-\alpha)/\beta^2 \end{cases}, \tag{19}$$

or inversely

$$\begin{cases} \alpha = (\sigma^2 - m^2)/\sigma^2 \\ \beta = m/\sigma^2, \end{cases}, \tag{20}$$

so that we can use these relations to determine $m$ and $\sigma^2$. Note also that the corresponding entropy of the final result is a byproduct of the function. Table (1) gives some numerical results obtained by ME_DENS1 program (See Annex).

| | | | Table 1. | | |
|---|---|---|---|---|---|
| $\mu_1$ | $\mu_2$ | $\alpha$ | $\beta$ | $m$ | $\sigma^2$ |
| 0.2000 | -3.0000 | 0.2156 | -3.0962 | 0.2533 | 0.0818 |
| 0.2000 | -2.0000 | -0.4124 | -6.9968 | 0.2019 | 0.0289 |
| 0.3000 | -1.5000 | -0.6969 | -5.3493 | 0.3172 | 0.0593 |

The next example is the case of a quartic distribution

$$p(x) = \exp\left[-\sum_{n=0}^{4} \lambda_n x^n\right]. \tag{21}$$

This distribution can be considered as a ME distribution when the constraints are

$$\mathrm{E}\{x^n\} = \int x^n p(x)\,\mathrm{d}x = \mu_n, \quad n = 0,\ldots,4 \quad \text{with} \quad \mu_0 = 1. \tag{22}$$

Now consider the following problem : Given $\mu_n, n = 1,\ldots,4$ calculate $\lambda_n, n = 0,\ldots,4$. This can be done by the ME_DENS2 program. Table (2) gives some numerical results obtained by this program:

| Table 2. | | | | | | | | |
|---|---|---|---|---|---|---|---|---|
| $\mu_1$ | $\mu_2$ | $\mu_3$ | $\mu_4$ | $\lambda_0$ | $\lambda_1$ | $\lambda_2$ | $\lambda_3$ | $\lambda_4$ |
| 0 | 0.2 | 0.05 | 0.10 | 0.1992 | 1.7599 | 2.2229 | -3.9375 | 0.4201 |
| 0 | 0.3 | 0.00 | 0.15 | 0.9392 | 0.000 | -3.3414 | 0.0000 | 4.6875 |
| 0 | 0.3 | 0.00 | 0.15 | 0.9392 | 0.000 | -3.3414 | 0.0000 | 4.6875 |

These examples show how to use the proposed programs. A third example is also given in Annex which shows how to use the ME_DENS3 program which considers the case of trigonometric moments.

## CONCLUSIONS

In this paper we addressed first the class of ME distributions when the available data are a finite set of expectations $\mu_n = \mathrm{E}\{\phi_n(x)\}$ of some known functions $\phi_n(x), n = 0, \ldots, N$. We proposed then three Matlab programs to solve this problem by a Newton–Raphson method in general case, in case of geometrical moments data where $\phi_n(x) = x^n$ and in case of trigonometrical moments where $\phi_n(x) = \exp(-jn\omega_0 x)$. Finally, we gave some numerical results for some special examples who show how to use the proposed programs.

# ANNEX A

```
1  function [lambda,p,entr]=me_dens1(mu,x,lambda0)
2  %ME_DENS1
3  %  [LAMBDA,P,ENTR]=ME_DENS1(MU,X,LAMBDA0)
4  %  This program calculates the Lagrange Multipliers of the ME
5  %  probability density functions p(x) from the knowledge of the
6  %  N contstraints in the form:
7  %  E{fin(x)}=MU(n)   n=0:N    with fi0(x)=1, MU(0)=1.
8  %
9  %  MU     is a table containing the constraints MU(n),n=1:N.
10 %  X      is a table defining the range of the variation of x.
11 %  LAMBDA0 is a table containing the first estimate of the LAMBDAs.
12 %         (This argument is optional.)
13 %  LAMBDA is a table containing the resulting Lagrange parameters.
14 %  P      is a table containing the resulting pdf p(x).
15 %  ENTR   is a table containing the entropy values at each
16 %         iteration.
17 %
18 %  Author: A. Mohammad-Djafari
19 %  Date   : 10-01-1991
20 %
21 mu=mu(:); mu=[1;mu];           % add mu(0)=1
22 x=x(:); lx=length(x);          % x axis
23 xmin=x(1); xmax=x(lx); dx=x(2)-x(1);
24 %
25 if(nargin == 2)                % initialize LAMBDA
26   lambda=zeros(size(mu));              % This produces a uniform
27   lambda(1)=log(xmax-xmin);    % distribution.
28 else
29   lambda=lambda0(:);
30 end
31 N=length(lambda);
32 %
33 fin=fin1_x(x);                 % fin1_x(x) is an external
34 %                              % function which provides fin(x).
35 iter=0;
36 while 1                        % start iterations
37   iter=iter+1;
38   disp('---------------'); disp(['iter=',num2str(iter)]);
39 %
40   p=exp(-(fin*lambda));        % Calculate p(x)
41   plot(x,p);                   % plot it
42 %
43   G=zeros(N,1);                % Calculate Gn
44   for n=1:N
45    G(n)=dx*sum(fin(:,n).*p);
46   end
47 %
48   entr(iter)=lambda'*G(1:N);   % Calculate the entropy value
49   disp(['Entropy=',num2str(entr(iter))])
50 %
51   gnk=zeros(N,N);              % Calculate gnk
52   gnk(1,:)=-G'; gnk(:,1)=-G;   % first line and first column
53   for i=2:N                    % lower triangle part of the
54    for j=2:i                   % matrix G
55     gnk(i,j)=-dx*sum(fin(:,j).*fin(:,i).*p);
56    end
57   end
58   for i=2:N                    % uper triangle part of the
59    for j=i+1:N                 % matrix G
60     gnk(i,j)=gnk(j,i);
61    end
62   end
63 %
64   v=mu-G;                      % Calculate v
65   delta=gnk\v;                 % Calculate delta
66   lambda=lambda+delta;         % Calculate lambda
67   eps=1e-6;                    % Stopping rules
68   if(abs(delta./lambda)<eps),                     break, end
69   if(iter>2)
70    if(abs((entr(iter)-entr(iter-1))/entr(iter))<eps),break, end
71   end
72 end
73 %
74 p=exp(-(fin*lambda));          % Calculate the final p(x)
75 plot(x,p);                     % plot it
76 entr=entr(:);
77 disp('-----  END  -------')
```

```
 1  %---------------------------------
 2  %ME1
 3  %  This script shows how to use the function ME_DENS1
 4  %  in the case of the Gamma distribution. (see Example 1.)
 5  xmin=0.0001; xmax=1; dx=0.01;   % define the x axis
 6  x=[xmin:dx:xmax]';
 7  mu=[0.3,-1.5]';                 % define the mu values
 8  [lambda,p,entr]=me_dens1(mu,x);
 9  alpha=-lambda(3);   beta=lambda(2);
10  m=(1+alpha)/beta; sigma=m/beta;
11  disp([mu' alpha beta m sigma entr(length(entr))])
12  %---------------------------------
```

```
 1  function fin=fin1_x(x);
 2  % This is the external function which calculates
 3  % the fin(x) in the special case of the Gamma distribution.
 4  % This is to be used with ME_dens1.
 5    M=3;
 6    fin=zeros(length(x),M);
 7    fin(:,1)=ones(size(x));
 8    fin(:,2)=x;
 9    fin(:,3)=log(x);
10  return
```

```
1  function [lambda,p,entr]=me_dens2(mu,x,lambda0)
2  %ME_DENS2
3  %   [LAMBDA,P,ENTR]=ME_DENS2(MU,X,LAMBDA0)
4  %   This program calculates the Lagrange Multipliers of the ME
5  %   probability density functions p(x) from the knowledge of the
6  %   N moment contstraints in the form:
7  %   E{x^n}=mu(n)    n=0:N    with  mu(0)=1.
8  %
9  %   MU     is a table containing the constraints MU(n),n=1:N.
10 %   X      is a table defining the range of the variation of x.
11 %   LAMBDA0 is a table containing the first estimate of the LAMBDAs.
12 %          (This argument is optional.)
13 %   LAMBDA is a table containing the resulting Lagrange parameters.
14 %   P      is a table containing the resulting pdf p(x).
15 %   ENTR   is a table containing the entropy values at each
16 %          iteration.
17 %
18 %   Author: A. Mohammad-Djafari
19 %   Date   : 10-01-1991
20 %
21 mu=mu(:); mu=[1;mu];            % add mu(0)=1
22 x=x(:); lx=length(x);           % x axis
23 xmin=x(1); xmax=x(lx); dx=x(2)-x(1);
24 %
25 if(nargin == 2)                 % initialize LAMBDA
26   lambda=zeros(size(mu));                % This produces a uniform
27   lambda(1)=log(xmax-xmin);     % distribution.
28 else
29   lambda=lambda0(:);
30 end
31 N=length(lambda);
32 %
33 M=2*N-1;                        % Calcul de fin(x)=x.^n
34 fin=zeros(length(x),M);         %
35 fin(:,1)=ones(size(x));                % fi0(x)=1
36 for n=2:M
37  fin(:,n)=x.*fin(:,n-1);
38 end
39 %
40 iter=0;
41 while 1                         % start iterations
42   iter=iter+1;
43   disp('---------------'); disp(['iter=',num2str(iter)]);
44 %
45   p=exp(-(fin(:,1:N)*lambda));  % Calculate p(x)
46   plot(x,p);                    % plot it
47 %
48   G=zeros(M,1);                 % Calculate Gn
49   for n=1:M
50    G(n)=dx*sum(fin(:,n).*p);
51   end
52 %
53   entr(iter)=lambda'*G(1:N);    % Calculate the entropy value
54   disp(['Entropy=',num2str(entr(iter))])
55 %
56   gnk=zeros(N,N);               % Calculate gnk
57   for i=1:N                     % Matrix G is a Hankel matrix
58    gnk(:,i)=-G(i:N+i-1);
59   end
60 %
61   v=mu-G(1:N);                  % Calculate v
62   delta=gnk\v;                  % Calculate delta
63   lambda=lambda+delta;          % Calculate lambda
64   eps=1e-6;                     % Stopping rules
65   if(abs(delta./lambda)<eps),                            break, end
66   if(iter>2)
67    if(abs((entr(iter)-entr(iter-1))/entr(iter))<eps,break, end
68   end
69 end
70 %
71 p=exp(-(fin(:,1:N)*lambda));    % Calculate the final p(x)
72 plot(x,p);                      % plot it
73 entr=entr(:);
74 disp('-----  END  -------')
75 end

1 %ME2
2 %  This script shows how to use the function ME_DENS2
3 %  in the case of the quartic distribution. (see Example 2.)
4 xmin=-1; xmax=1; dx=0.01;          % define the x axis
5 x=[xmin:dx:xmax]';
6 mu=[0.1,.3,0.1,.15]';              % define the mu values
7 [lambda,p,entr]=me_dens2(mu,x);
8 disp([mu;lambda;entr(length(entr))]')
```

```matlab
 1  function [lambda,p,entr]=me_dens3(mu,x,lambda0)
 2  %ME_DENS3
 3  %   [LAMBDA,P,ENTR]=ME_DENS3(MU,X,LAMBDA0)
 4  %   This program calculates the Lagrange Multipliers of the ME
 5  %   probability density functions p(x) from the knowledge of the
 6  %   Fourier moments values :
 7  %   E{exp[-j n w0 x]}=mu(n)    n=0:N   with  mu(0)=1.
 8  %
 9  %   MU     is a table containing the constraints MU(n),n=1:N.
10  %   X      is a table defining the range of the variation of x.
11  %   LAMBDA0 is a table containing the first estimate of the LAMBDAs.
12  %          (This argument is optional.)
13  %   LAMBDA is a table containing the resulting Lagrange parameters.
14  %   P      is a table containing the resulting pdf p(x).
15  %   ENTR   is a table containing the entropy values at each
16  %          iteration.
17  %
18  % Author: A. Mohammad-Djafari
19  % Date  : 10-01-1991
20  %
21  mu=mu(:);mu=[1;mu];            % add mu(0)=1
22  x=x(:); lx=length(x);          % x axis
23  xmin=x(1); xmax=x(lx); dx=x(2)-x(1);
24  if(nargin == 2)                % initialize LAMBDA
25    lambda=zeros(size(mu));           % This produces a uniform
26    lambda(1)=log(xmax-xmin);    % distribution.
27  else
28    lambda=lambda0(:);
29  end
30  N=length(lambda);
31  %
32  M=2*N-1;                       % Calculate fin(x)=exp[-jnw0x]
33  fin=fin3_x(x,M);               % fin3_x(x) is an external
34  %                              % function which provides fin(x).
35  iter=0;
36  while 1                        % start iterations
37    iter=iter+1;
38    disp('---------------'); disp(['iter=',num2str(iter)]);
39  %
40                                 % Calculate p(x)
41    p=exp(-real(fin(:,1:N))*real(lambda)+imag(fin(:,1:N))*imag(lambda));
42    plot(x,p);                   % plot it
43  %
44    G=zeros(M,1);                % Calculate Gn
45    for n=1:M
46     G(n)=dx*sum(fin(:,n).*p);
47    end
48    %plot([real(G(1:N)),real(mu),imag(G(1:N)),imag(mu)])
49  %
50    entr(iter)=lambda'*G(1:N);   % Calculate the entropy
51    disp(['Entropy=',num2str(entr(iter))])
52  %
53    gnk=zeros(N,N);              % Calculate gnk
54    for n=1:N                    % Matrix gnk is a Hermitian
55     for k=1:n                   % Toeplitz matrix.
56      gnk(n,k)=-G(n-k+1);        % Lower triangle part
57     end
58    end
59    for n=1:N
60     for k=n+1:N
61      gnk(n,k)=-conj(G(k-n+1));  % Upper triangle part
62     end
63    end
64  %
65    v=mu-G(1:N);                 % Calculate v
66    delta=gnk\v;                 % Calculate delta
67    lambda=lambda+delta;         % Calculate lambda
68    eps=1e-3;                    % Stopping rules
69    if(abs(delta)./abs(lambda)<eps),              break, end
70    if(iter>2)
71     if(abs((entr(iter)-entr(iter-1))/entr(iter))<eps),break, end
72    end
73  end
74                                 % Calculate p(x)
75  p=exp(-real(fin(:,1:N))*real(lambda)+imag(fin(:,1:N))*imag(lambda));
76  plot(x,p);                     % plot it
77  entr=entr(:);
78  disp('-----  END  -------')
79  end
```

```
 1  %ME3
 2  %  This scripts shows how to use the function ME_DENS3
 3  %  in the case of the trigonometric moments.
 4  clear;clf
 5  xmin=-5; xmax=5; dx=0.5;          % define the x axis
 6  x=[xmin:dx:xmax]';lx=length(x);
 7  p=(1/sqrt(2*pi))*exp(-.5*(x.*x));% Gaussian distribution
 8  plot(x,p);title('p(x)')
 9  %
10  M=3;fin=fin3_x(x,M);              % Calculate fin(x)
11  %
12  mu=zeros(M,1);                    % Calculate mun
13  for n=1:M
14    mu(n)=dx*sum(fin(:,n).*p);
15  end
16  %
17  w0=2*pi/(xmax-xmin);w=w0*[0:M-1]'; % Define the w axis
18  %
19  mu=mu(2:M);                       % Attention : mu(0) is added
20                                    % in ME_DENS3
21  [lambda,p,entr]=me_dens3(mu,x);
22  disp([mu;lambda;entr(length(entr))]')
```

```
 1  function fin=fin3_x(x,M);
 2  % This is the external function which calculates
 3  % the fin(x) in the special case of the Fourier moments.
 4  % This is to be used with ME_DENS3.
 5  %
 6  x=x(:); lx=length(x);             % x axis
 7  xmin=x(1); xmax=x(lx); dx=x(2)-x(1);
 8  %
 9  fin=zeros(lx,M);                  %
10  fin(:,1)=ones(size(x));           % fi0(x)=1
11  w0=2*pi/(xmax-xmin);jw0x=(sqrt(-1)*w0)*x;
12  for n=2:M
13    fin(:,n)=exp(-(n-1)*jw0x);
14  end
15  return
```